\documentstyle[epsfig]{mn}

\def\be{\begin{equation}}
\def\ee{\end{equation}}

\def\HI{\hbox{H$\scriptstyle\rm I\ $}}

\def\HeI{\hbox{He$\scriptstyle\rm I\ $}}
\def\HeII{\hbox{He$\scriptstyle\rm II\ $}}

\def\gtsima{$\; \buildrel > \over \sim \;$}
\def\ltsima{$\; \buildrel < \over \sim \;$}
\def\prosima{$\; \buildrel \propto \over \sim \;$}
\def\gsim{\lower.5ex\hbox{\gtsima}}
\def\lsim{\lower.5ex\hbox{\ltsima}}
\def\simgt{\lower.5ex\hbox{\gtsima}}
\def\simlt{\lower.5ex\hbox{\ltsima}}
\def\simpr{\lower.5ex\hbox{\prosima}}

\begin{document}
\title[Early Reionization by the First Galaxies]
{Early Reionization by the First Galaxies}

\author[B. Ciardi et al.]
{B. Ciardi$^{1}$, A. Ferrara$^{2}$ \&  S. D. M. White$^{1}$\\ 
$^1$ Max-Planck-Institut f\"{u}r Astrophysik, 85741 Garching, Germany\\
$^2$ SISSA/International School for Advanced Studies, Via Beirut 4, 34014 Trieste, Italy\\}

\maketitle
\vspace {7cm}

\begin{abstract}
Large-scale polarization of the cosmic microwave background measured
by the {\it WMAP} satellite requires a mean optical depth to Thomson
scattering, $\tau_e \sim 0.17$. 
The reionization of the universe must
therefore have begun at relatively high redshift.
We have studied the
reionization process using supercomputer simulations of a large and
representative region of a universe which has cosmological parameters
consistent with the {\it WMAP} results. Our simulations follow both
the radiative transfer of ionizing photons and the formation and
evolution of the galaxy population which produces them. A previously
published model with a standard stellar IMF, with a moderate photon
escape fraction from galaxies and with ionizing photon production as
expected for zero metallicity stars produces $\tau_e = 0.104$, which is
within 1.0 to $1.5\sigma$ of the ``best'' {\it WMAP} value. Values up
to 0.16 can be produced by taking larger escape fractions or a top
heavy IMF. The data do not require a separate populations of
``miniquasars'' or of stars forming in objects with total masses below
$10^9$~M$_{\odot}$.  Reconciling such early reionization with the
observed Gunn-Peterson troughs in $z>6$ quasars may be challenging.
Possible resolutions of this problem are discussed.
\end{abstract}


\section{Motivation}
It has recently become possible to study the reionization of the
universe in some detail. Massive numerical computations can now follow
not only the clustering of dark matter and gas, but also the formation
of galaxies and the propagation of ionizing photons in a highly
inhomogeneous, partially ionized, intergalactic medium (IGM). In spite
of many poorly understood details concerning the physics of star
formation, and the approximations inherent in the various numerical
treatments of radiative transfer (for a recent review see Maselli,
Ferrara \& Ciardi 2003), a number of independent studies have
converged on a relatively late ($z_r \simlt 8$ to 10) epoch for
complete reionization of the IGM within current ``concordance''
(i.e. flat, $\Lambda$-dominated) cosmological models (e.g. Ciardi et
al. 2000, hereafter CFGJ; Gnedin 2000; Razoumov et al. 2002; Ciardi,
Stoehr \& White 2003, hereafter CSW).  

This conclusion appears challenged by results from the {\it WMAP}
satellite (Kogut et al. 2003; Spergel et al. 2003).  This experiment
has detected an excess in the CMB TE cross-power spectrum on large
angular scales ($\ell < 7$) indicating an optical depth to the CMB
last scattering surface of $\tau_e=0.16$. The uncertainty quoted for
this number depends on the analysis technique employed. Fitting the TE
cross power spectrum to $\Lambda$CDM models in which all parameters
except $\tau_e$ take their best fit values based on the TT power
spectrum, Kogut et al. (2003) obtain a 68\% confidence range,
$0.13<\tau_e<0.21$. Fitting all parameters simultaneously to the TT+TE
data, Spergel et al. (2003) obtain $0.095<\tau_e< 0.24$.  Including
additional data external to {\it WMAP}, these authors were able to
shrink their confidence interval to $0.11<\tau_e<0.23$.  Finally, by
assuming that the {\it observed} TT power spectrum is scattered to
produce the observed TE cross-power spectrum Kogut et al. (2003) infer
$0.12<\tau_e<0.20$. It is unclear which of these uncertainty estimates
to prefer. Most $\tau_e$ values in these ranges require a substantial
fraction of the universe to be ionized before redshift 10. For
example, $0.12<\tau_e<0.20$ translates into $13 < z_r < 19$
in a model where reionization is instantaneous.

Apparently, reionization occurred earlier than expected. Is this
discrepancy real? In practice, recovering the ``reionization epoch''
is subject to some ambiguity related to the specific reionization
history assumed (see e.g. Bruscoli, Scannapieco \& Ferrara 2002). What
might reionization models have overlooked?  The discrepancy is not
dependent on the particular cosmological parameters adopted, since
{\it WMAP} has confirmed the previous concordance model.  Hence the
``oversight'' must be of astrophysical nature.  Several effects might
have produced rapid early evolution: a contribution from low-mass
pregalactic systems (in the jargon, Population III objects);
unexpectedly high star formation efficiency; unexpectedly high
ionizing photon production; an unexpectedly large probability for
ionizing photons to escape into the IGM; a possible population of
early ``miniquasars''.  In this paper, we point out that efficient
production and escape of ionizing photons is sufficient to account for
the data within conventional galaxy formation models. The data do not
{\it require} very massive stars or miniquasars, although the high
efficiencies needed may point to near-zero metallicities or to a
top-heavy stellar Initial Mass Function (IMF) at early times.

\section{Simulating Cosmic Reionization}

\begin{figure*}
\psfig{figure=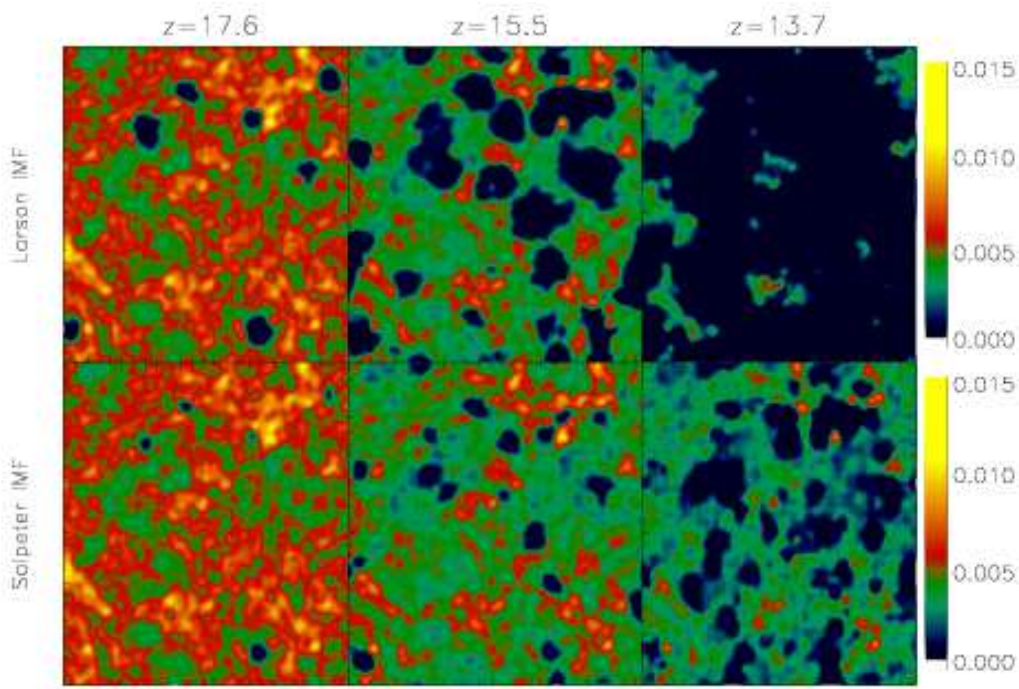,height=10cm}
\caption{\label{fig01}\footnotesize{Slices through the simulation
boxes. The six panels show the neutral hydrogen number density for the
L20 (upper panels) and the S20 (lower panels) runs, at redshifts, from
left to right, $z=17.6, 15.5$~and 13.7. The box for the radiative
transfer simulation has a comoving length of $L=20 h^{-1}$~Mpc.}}
\end{figure*}

In CSW we studied cosmic reionization through a combination of
high-resolution N-body simulations (to describe the distribution of
dark matter and diffuse gas), a semi-analytic model of galaxy
formation (to track the sources of ionization) and the Monte Carlo
radiative transfer code {\tt CRASH} (to follow the propagation of
ionizing photons in the IGM; Ciardi et al. 2001, CFMR; Maselli,
Ferrara \& Ciardi 2003). Here, we briefly summarize the features of
the above approach which are relevant to the present study.

The simulations are based on a $\Lambda$CDM ``concordance'' cosmology
with $\Omega_m$=0.3, $\Omega_{\Lambda}$=0.7, $h=$0.7, $\Omega_b$=0.04,
$n$=1 and $\sigma_8$=0.9. These parameters are within the {\it WMAP}
experimental error bars (Spergel et al. 2003).  The re-simulation
technique described in Springel et al. (2001, hereafter SWTK) and the
N-body code {\tt GADGET} (Springel, Yoshida \& White 2001) were used
to follow at high resolution the dark matter distribution within an
approximately spherical subregion of diameter about $50 h^{-1}$~Mpc
within a much larger cosmological volume. Within this subregion a cube
of comoving side $L=20 h^{-1}$~Mpc was used for the detailed radiative
transfer modeling.  The location and mass of dark matter halos was
determined with a friends-of-friends algorithm.  Gravitationally bound
substructures were identified within the halos with the algorithm
{\tt SUBFIND} (SWTK) and were used to build the merging tree for halos
and subhalos following the prescription of SWTK. The smallest resolved
halos have masses of $M \simeq 10^9$~M$_\odot$ and they start forming
in our box at $z\sim 20$.  We model the galaxy population via the
semi-analytic technique of Kauffmann et al. (1999) as implemented by
SWTK.  At the end of this process we obtain a mock catalogue of
galaxies for each of the 51 simulation outputs, containing for each
galaxy, among other quantities, its position, stellar mass and star
formation rate.  The simulations match the observed global star
formation rate density evolution for $z \le 5$. The N-body simulation
used in this paper is the M3 simulation of CSW and all galaxy
formation modeling is identical to that in the earlier paper.

Reionization simulations require a mass resolution high enough to
follow the formation and evolution of the objects producing the bulk
of the ionizing radiation. At the same time, large simulation volumes must
be considered to avoid biases due to cosmic variance on small scales.
Moreover, a reliable treatment of the radiative transfer of ionizing
photons in the IGM is needed. So far, although several numerical
approaches have been proposed (e.g. Gnedin \& Ostriker 1997; CFGJ;
Chiu \& Ostriker 2000; Gnedin 2000; Razoumov et al. 2002) only the
simulations described in CSW fulfill all the above requirements (refer
to that paper for an extensive discussion).

One of the aims of the present study is to assess the influence of the
IMF on reionization. We have thus inferred the emission properties of
our model galaxies by assuming a time-dependent spectrum of a simple
stellar population of metal-free stars in the mass range up to
40~M$_\odot$, with two different IMFs (CFMR): (i) a Salpeter IMF and
(ii) a Larson IMF, i.e. a Salpeter function at the upper mass end
which falls off exponentially below a characteristic stellar mass,
$M_c=5$~M$_\odot$.  The Larson IMF has a specific photon flux at the
Lyman continuum which is about 4 times the Salpeter one. The total
number of ionizing photons per solar mass for a Larson IMF is
$\approx 4 \times 10^{60}$~M$_\odot^{-1}$.

Of the emitted ionizing photons, only a fraction $f_{esc}$ will
actually be able to escape into the IGM.  This quantity is poorly
determined both theoretically and observationally: actually, $f_{esc}$
may well vary with, e.g., redshift, mass and structure of a galaxy, as
well as with the ionizing photon production rate (e.g. Wood \& Loeb
1999; Ricotti \& Shull 2000; Ciardi, Bianchi \& Ferrara 2002).
However, at moderate redshift, recent results on the
opacity evolution of the Ly$\alpha$ forest (Bianchi, Cristiani \& Kim
2001) constrain $f_{esc}$ to be smaller than 20\% in order not to
over-produce the cosmic UV background. Although this limit may not apply
to the high redshift universe, we consider this
value a physically plausible upper limit.

\begin{table}
\centerline{\begin{tabular}[t] {|l|l|l|l|r|} \hline
{\em RUN} & {\em IMF} & {\em $f_{esc}$}\\ \hline
S5  & Salpeter & 5\% \\
S20 & Salpeter & 20\%\\
L20 & Larson   & 20\%\\ \hline
\end{tabular}}
\caption{Parameters of the simulations: Initial Mass Function, IMF; photon
escape fraction, $f_{esc}$.}
\label{tab}
\end{table}

Given these assumptions, the simulations of reionization described
above have been run for the three different parameter combinations in
Table~\ref{tab}.  Run S5 (extensively described in CSW as the M3
case), yields the lowest IGM ionization power input, whereas L20
maximizes it; run S20 is an intermediate case. The critical parameter
differentiating these runs is the number of ionizing photons escaping
a galaxy into the IGM for each solar mass of long-lived stars which it
forms. Each simulation can be thought of as corresponding to models
with varying IMF, stellar luminosity beyond the Lyman limit and
value of $f_{esc}$, provided this quantity is kept fixed (see CSW).

We quantify the agreement between our simulations and the {\it WMAP} data
primarily through the optical depth to electron scattering, $\tau_e$, given by:
\begin{equation}
\tau_e(z)=\int_0^z \sigma_T n_e(z') c \left\vert \frac{dt} {dz'} \right\vert dz',
\label{tau}
\end{equation}
where $\sigma_T=6.65 \times 10^{-25}$~cm$^2$ is the Thomson cross
section and $n_e(z')$ is the mean electron number density at $z'$.
We will compare our simulated $\tau_e$ with the ``model independent''
estimate of Kogut et al. (2003) $\tau_e = 0.16\pm 0.04$ (68\%
confidence range), but it should be borne in mind that the Spergel
et al. (2003) joint analysis of {\it WMAP} and external data suggests
a significantly larger allowed range. 

\section{Results}

In order to clarify the differences between our runs, we start from a
visual inspection of simulated maps. Fig.~\ref{fig01} shows the redshift
evolution of the \HI number density for the L20 (upper panels) and the
S20 (lower) runs (illustrative maps for the S5 runs can be found in
CSW).  Highly ionized regions (dark areas) are produced by the young
galaxies in the box and are well resolved in the maps. They initially
occupy a small fraction of the volume, and are typically larger in the
L20 run because of the higher ionizing power of the sources.  Their
shape, particularly for larger ones, appears distorted by nearby high
density peaks. (To a good approximation these correspond to peaks in
the \HI distribution.) The ionization front slows when it encounter
such overdensities because of their higher recombination rate. By
redshift $z=15.5$ several bubbles are close to overlap in the L20 run
(see upper-right corner of the central panel) whereas in the S20 run
the filling factor is still small. Finally, by $z=13.7$ the
overlapping fronts have cleared out most of the volume in L20 with
tiny \HI islands surviving thanks to their high density; reionization
in the S20 run, on the other hand, is far from complete.  

These remarks can be complemented by the more quantitative analysis
shown in Fig.~2, which plots the (normalized) distribution function of
the simulated electron density at various redshifts for the two
extreme runs, L20 and S5. This distribution is defined as the fraction
of the total volume filled with gas with free electron density in each
logarithmic bin of $n_e$. Its bimodal shape reflects the two-phase
(neutral + ionized) structure of the IGM.  The mean particle density
in the redshift range $13.3< z < 18.5$ is around $10^{-3}$~cm$^{-3}$,
so the rightmost peak has $n_e \approx n$ and corresponds to the
ionized phase.  Conversely, the left peak is associated with mostly
neutral gas. The general evolutionary trend is a continuous transfer
of matter from the left peak to the right one as redshift
decreases. By $z=13.3$ we find that the mass-averaged neutral hydrogen
fraction is only 2\% for the L20 model, whereas for the S5 and S20
cases it is still 89\% and 63\% respectively.

The redshift evolution of the volume-averaged ionization
fraction, $x_v$, essentially coincident with mass-averaged curve
(CSW), is reported for each of our three runs in the inset 
of Fig.~3. An L20 vs S20 comparison shows that the former typically has
an $x_v$ value $\approx$~3 times higher, reaching
complete ionization ($x_v \approx 1$) at $z_r \approx 13$. Run S20, 
on the other hand, reaches complete reionization at $z_r \approx 11$. 
In run S5 reionization proceeds much more gradually and it is only 
at $z_r \approx 8$ that the value $x_v \approx 1$ is reached.

Finally, we have calculated the evolution of $\tau_e$
(Fig.~\ref{fig03}) corresponding to the above reionization histories
as follows. Prior to complete reionization, $n_e(z)$ in eq.~\ref{tau}
is obtained from the simulations; after the reionization epoch, we
simply assume complete H and \HeI ionization throughout
the box. We also assume \HeII reionization at $z=3$.  The three runs
yield the values $\tau_e=0.104$ (S5), $\tau_e=0.132$ (S20) and
$\tau_e=0.161$ (L20).  A value $\tau_e=0.16$ is also obtained if one
assumes instantaneous reionization at $z_r \approx 16$ (dotted line),
i.e. three redshift units higher than the actual epoch of complete
reionization in the model. At $z=16$ the ionization fraction
is $x_v \approx 0.3$ in L20.

\begin{figure}
\psfig{figure=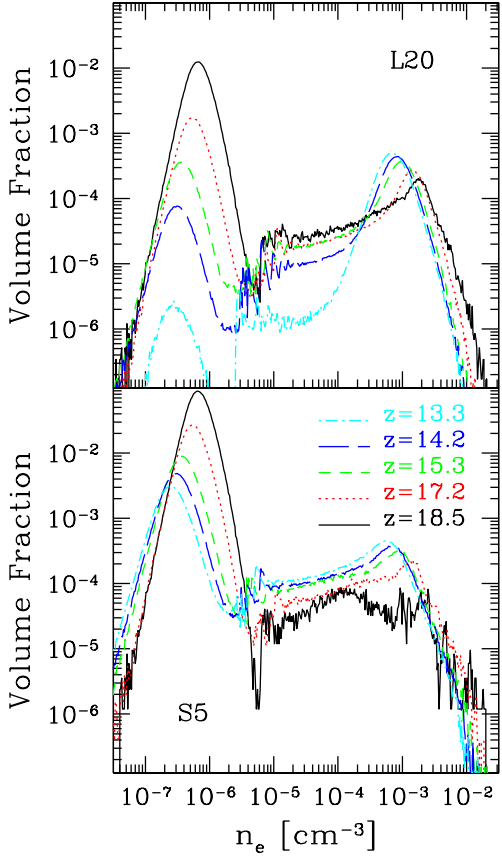,height=12cm}
\caption{\label{fig02}\footnotesize{Evolution of the (normalized) electron density
distribution function, $n_e$, for the L20 (upper panel) and S5 (lower) runs. Curves refer
to different redshifts, as in the label.}} 
\end{figure}
\begin{figure}
\psfig{figure=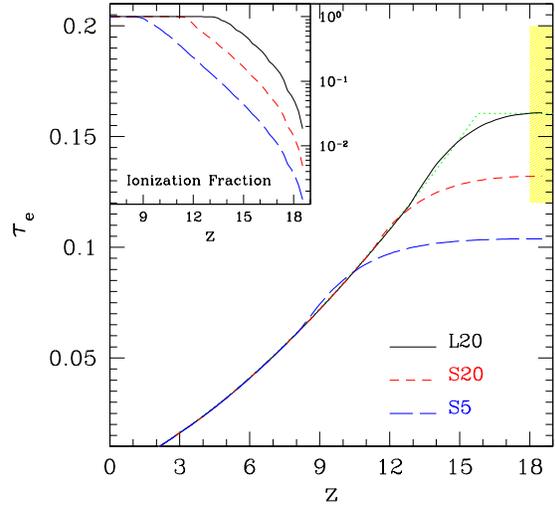,height=7.5cm}
\caption{\label{fig03}\footnotesize{Redshift evolution of the electron
optical depth, $\tau_e$, for the S5 (long-dashed line), S20 (short-dashed) and
L20 (solid) runs.  The dotted line refers to sudden reionization at $z
= 16$.  The shaded region indicates the optical depth $\tau_e = 0.16\pm
0.04$ (68\% CL) implied by the Kogut et al. (2003) ``model
independent'' analysis.  In the inset
the redshift evolution of the volume-averaged ionization fraction,
$x_v$, is shown for the three runs.}}
\end{figure}

\section{Discussion}

Our supercomputer simulations of galaxy formation and of the
propagation of ionization fronts have shown that recent {\it WMAP}
measurements of the optical depth to electron scattering
($\tau_e$=0.16$\pm 0.04$ according to the ``model independent''
analysis of Kogut et al.  (2003) or $\tau_e$=0.17$\pm 0.06$ according
to the parameter fits of Spergel et al. (2003)) are easily reproduced
by a model in which reionization is caused by the first stars in
galaxies with total masses of a few $\times 10^9$M$_\odot$. In order
to get sufficient early ionization, this phase of star formation must
supply a relatively high number of ionizing photons to the IGM for
each solar mass of long-lived stars which is formed. This requires
some combination of a high photon escape fraction, a top-heavy
IMF, and a high stellar production rate for ionizing photons, similar,
perhaps, to that typically inferred for metal-free stars. Among the
three models we have explored, the ``best'' {\it WMAP} value for
$\tau_e$ is matched assuming a moderately top-heavy IMF and an escape
fraction of 20\%. A Salpeter IMF with the same escape fraction gives
$\tau_e = 0.132$, which is still within all the suggested 68\%
confidence ranges. Decreasing $f_{esc}$ to 5\% gives $\tau_e =0.104$,
which disagrees with {\it WMAP} only at the 1.0 to $1.5\sigma$ level
depending on which uncertainty estimate one adopts. All these models
assume that ionizing photons are produced at the level expected for
metal-free stars.  It is clearly possible to reproduce the
experimental data without invoking exotica such as very massive
stars, early ``miniquasars'', or minihalos, i.e. halos cooled by
molecular hydrogen with virial temperatures $<10^4$~K.

In our best-fitting model (L20) reionization is essentially complete
by $z_r\approx 13$. This is difficult to reconcile with observations
of the Gunn-Peterson effect in $z>6$ quasars (Becker et al. 2001; Fan
et al. 2002). These imply a volume-averaged neutral fraction above
$10^{-3}$ and a mass-averaged neutral fraction $\sim 1$\% at
$z=6$. These values are reached in our L20 run at $z\approx 13$. Even for
our S5 model, they are reached before $z\approx 8$.  Thus the neutral
fraction at late times must be higher than in our models if the
Gunn-Peterson data are to be consistent with the {\it WMAP} findings.
A fascinating (although speculative) possibility is that the universe
was reionized twice (Cen 2002; Wyithe \& Loeb 2002) with a relatively
short redshift interval in which the IGM became neutral again.  The
maximum thickness of the neutral layer (ending at $z\sim 6$) for which
the L20 run remains consistent with the {\it WMAP} data, is $\Delta
z\sim 3$; this sets the end of the first reionization epoch at $z
\approx 9$. Distinguishing single- from two-epoched reionization 
models is in principle possible from the TE cross-correlation spectra
(Naselsky \& Chiang 2003); however, because of the high sensitivity
required, this measurement will have to await the {\it Planck}
mission.

What mechanism could have reduced the production of ionizing photons
enough to allow recombination at the end of the first reionization
epoch? Suppression of the galaxy formation process itself seems
implausible. An increase of the relevant filtering scale 
could have suppressed galaxy formation in halos below a typical
circular velocity of $30-40$~km~s$^{-1}$ (Gnedin 2000) but this is
well below the scale of the halos which host most of the star
formation in our models at redshifts below 10. Feedback by SNe might
reduce formation efficiencies (e.g. Madau, Ferrara \& Rees 2001;
Scannapieco et al. 2000) but seems likely to be more effective at
the high redshifts where we need efficient star formation than at the
later times when recombination is supposed to occur. A more plausible
possibility may be that galaxy formation continues apace, but that the
efficiency of ionizing photon production drops dramatically, perhaps
as a result of increasing stellar metallicity, of decreasing escape
fraction, or of changing IMF.

In many ways our models are quite conservative. We have simply
extrapolated conventional models for galaxy formation to higher
redshift assuming that feedback and star formation efficiencies 
can be scaled down to systems with mass a few $\times 10^9$M$_\odot$
and circular velocities $\sim 60$~km~s$^{-1}$ at $z\approx 15$. We have
adopted optimistic but not implausible values for ionizing photon
production efficiency and escape fraction. We do not require any stars
to form in low mass objects or through molecular cooling processes,
nor do we invoke any non-stellar sources of ionizing radiation.

Processes we have ignored may nonetheless play a significant role.  We
do not attempt to model the possible effects of (unresolved) minihalos.
Such objects rely on
hydrogen molecules for cooling, and it is unclear whether they can
form stars at all. As pointed out by CFGJ, and more recently by Wyithe
\& Loeb (2003), their contribution to the ionizing photon budget is in
any case expected to be negligible.  Minihalos could also increase the
IGM clumping factor, thus enhancing its recombination efficiency and
providing a potentially important sink for photons. If this effect
were strong, more extreme assumptions about the IMF and the escape
fractions would be required to ensure early reionization. One might
then be forced to invoke very massive stars.  For example, a 200~M$_\odot$
star, ending its life as a pair-instability supernova
(SN$_{\gamma\gamma}$), produces approximately $3\times 10^3
(Z/Z_\odot$) ionizing photons/H-atom.  Stars of this mass may cease to
form once the metallicity of the parent gas cloud exceeds $Z \approx
10^{-5\pm 1} Z_\odot$ (Bromm et al. 2001; Schneider et al. 2002). They
would then disappear after they have contributed about 0.003 - 0.3
photons/H-atom, hence prior to reionization. In addition, one could
posit stars outside the SN$_{\gamma\gamma}$ mass range, which end up
in very massive black holes. This may lead to a star formation
conundrum (Schneider et al. 2002): lacking the necessary heavy
element pollution, only extremely heavy stars would form indefinitely.
Although a mixture of the two populations, dominated by BH-progenitor
stars might be favored by observations of the near infrared background
(Salvaterra \& Ferrara 2002; Magliocchetti, Salvaterra \& Ferrara
2003), our current results suggest that the {\it WMAP} measurement of
$\tau_e$ does not require any such exotic objects.

\noindent
\section*{ACKNOWLEDGMENTS}
We thank R. Fabbri, F. Miniati, R. Salvaterra and R. Schneider for useful discussions
and in particular F. Stoehr for collaboration during the project.  
This work has been partially supported by the Research and Training Network
``The Physics of the Intergalactic Medium'' set up by the European
Community under the contract HPRN-CT-2000-00126.
AF acknowledges ESO hospitality through the Visiting Fellowship Program.

\end{document}